\documentclass[fleqn,10pt]{wlscirep}
\usepackage[utf8]{inputenc}
\usepackage[T1]{fontenc}
\title{Generative Machine Learning for Robust Free-Space Communication}

\author[1,+]{Sanjaya Lohani}
\author[1,*]{Ryan T. Glasser}
\affil[1]{Tulane University, New Orleans, LA 70118, USA}

\affil[+]{slohani@tulane.edu}

\affil[*]{rglasser@tulane.edu}

\begin{abstract}
Realistic free-space optical communications systems suffer from turbulent propagation of light through the atmosphere and detector noise at the receiver, which can significantly degrade the optical mode quality of the received state, increase cross-talk between modes, and correspondingly increase the symbol error ratio (SER) of the system.  In order to overcome these obstacles, we develop a state-of-the-art generative machine learning (GML) and convolutional neural network (CNN) system in combination, and demonstrate its efficacy in a free-space optical (FSO) communications setting.  The system corrects for the distortion effects due to turbulence and reduces detector noise, resulting in significantly lowered SERs and cross-talk at the output of the receiver, while requiring no feedback. This scheme is straightforward to scale, and may provide a concrete and cost effective technique to establishing long range classical and quantum communication links in the near future.
\end{abstract}
\begin{document}

\flushbottom
\maketitle
%
%
\thispagestyle{empty}
\section{Introduction}
\begin{figure}[h!] 
\centering\includegraphics[width=0.65\linewidth]{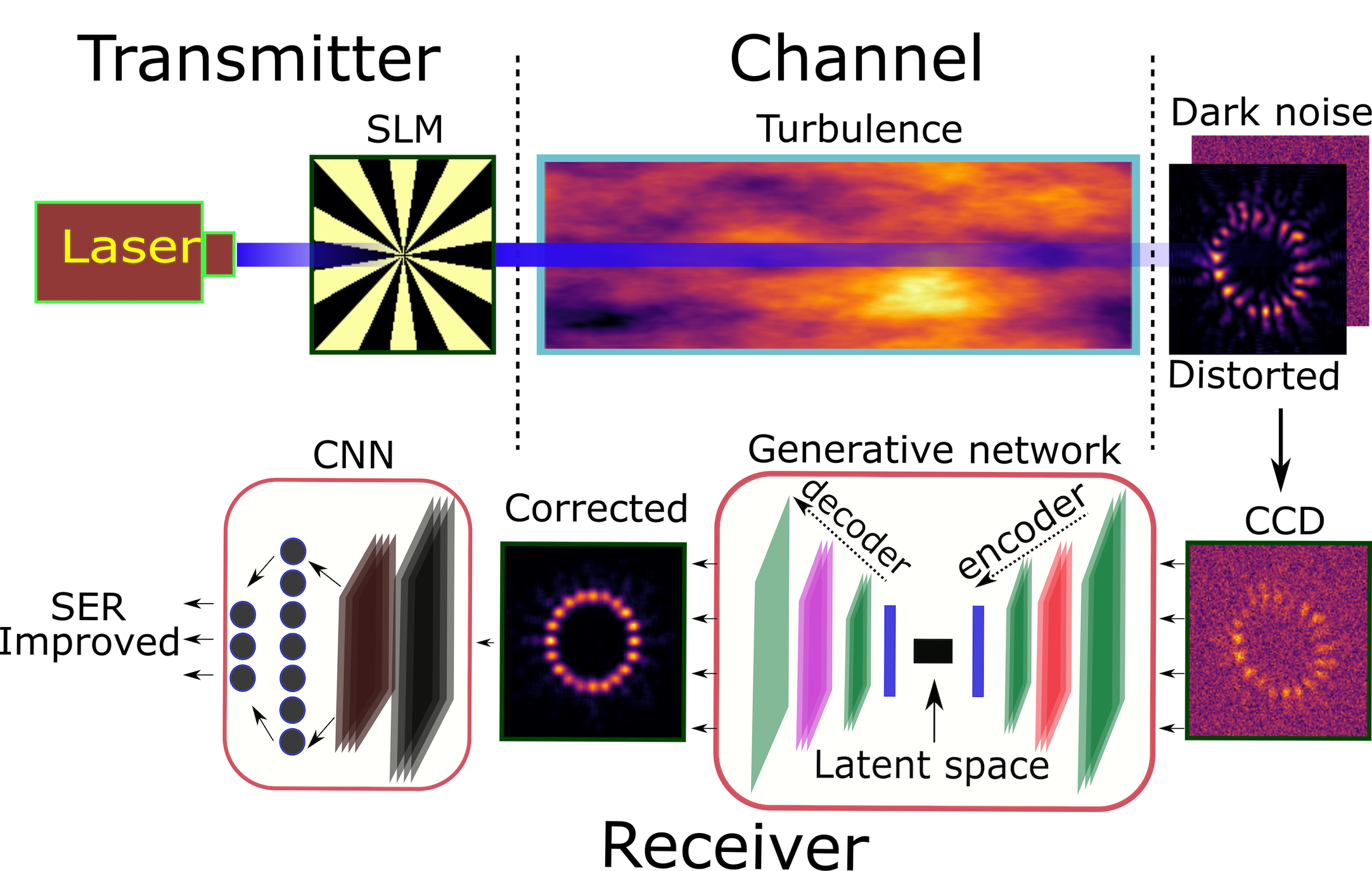}
\caption{Schematic of the robust free-space optical communications scheme.  Simulations include generating a desired optical mode, its propagation through turbulence, and added detector noise.  The resultant image at the receiver is then fed into the generative neural network, which generates a new, less distorted and noisy image.  This generated image is then classified by a convolutional neural network, allowing for the calculation of the SER and cross-talk between the modes.} 
\label{fig:Figure_1}
\end{figure}

\begin{figure}[h!] 
\centering\includegraphics[width=0.9\linewidth]{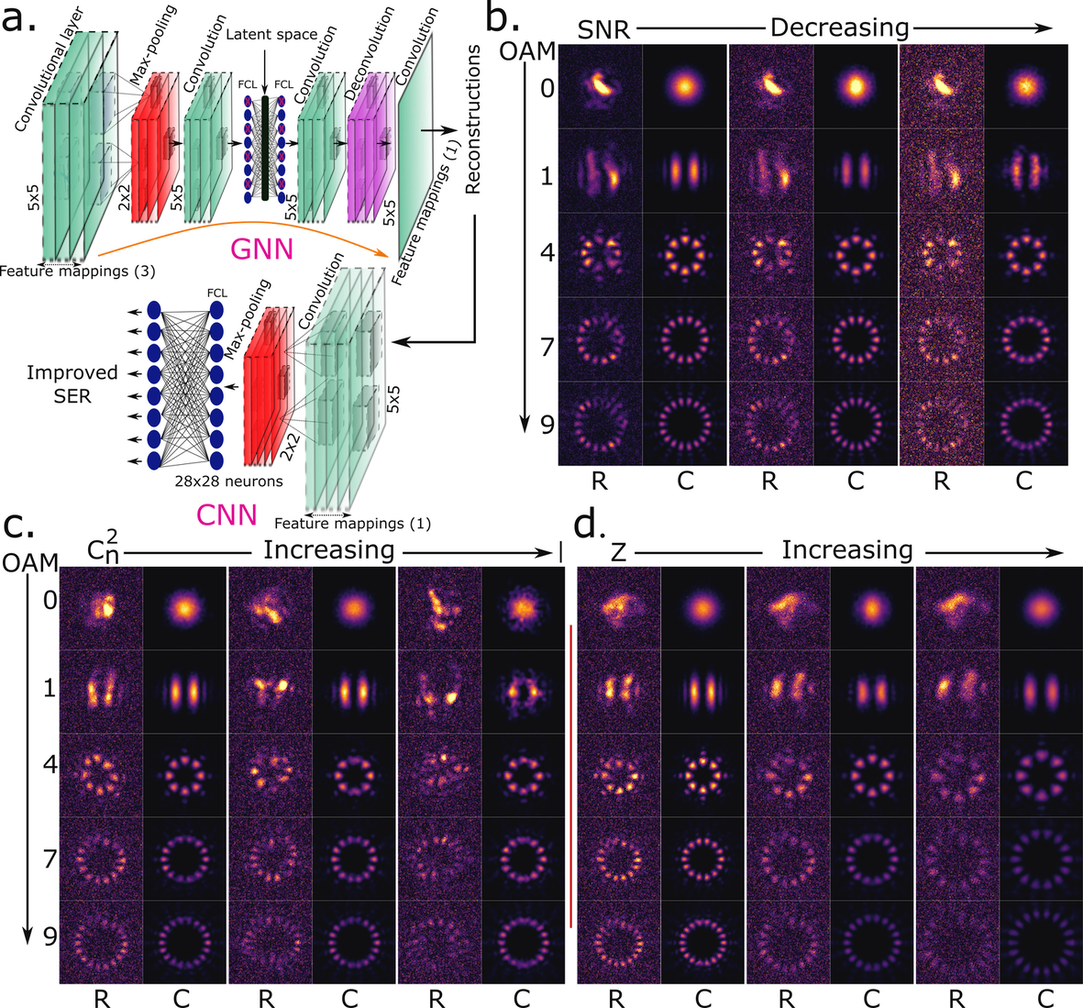}
\caption{(a) An architecture of neural networks consisting of a generative neural network (GNN) and a convolutional neural network (CNN). (b,c,d) Example images of various simulated optical modes that are distorted and noisy at the receiver (R), and the generated, corrected modes (C).  Decreasing the SNR (left: $-0.11$\,dB, middle: $-3.87$\,dB, right: $-5.91$\,dB) is shown in (b), increasing the turbulence strength $C_n^2$ (left: $C_n^2\,=\,5\times 10^{-14}\,m^{-2/3}$, middle: $C_n^2\,=\,7\times 10^{-14}\,m^{-2/3}$, right: $C_n^2\,=\,1\times 10^{-13}\,m^{-2/3}$) is shown in (c), and increasing the communications link distance ($Z$) (left: $400\,m$, middle: $600\,m$, right: $800\,m$) is shown in (d).  The superposition of OAM degree is increased in each row, from 0 (Gaussian) up to a value of $\pm9$.} 
\label{fig:Figure_2}
\end{figure}

The field of FSO communications provides an exciting route forward in wireless communication by making use of various multiplexing schemes, including frequency and wavelength division multiplexing, and more recently spatial multiplexing \cite{huang2014100,ren2015free,qu2017beyond,nejad2016orbital,milione20154,wang2012terabit}.  
A common method to implement the latter involves making use of orbital angular momentum (OAM), which is a degree-of-freedom that is in principle unbounded, thus allowing for the use of large alphabets in FSO communication links. For example, by generating and transmitting various superpositions of OAM states, which result in different ``petal pattern'' images in the spatial domain as shown in Figs.\,\ref{fig:Figure_1} and \ref{fig:Figure_2}, the alphabet size of the communications system may be significantly increased.  An integral aspect of the communications system, however, is the ability to effectively demodulate the signal at the receiver, which in this case corresponds to determining which OAM superposition was sent and received. In practice, realistic FSO communications systems involve the propagation of such signals through turbulence, and include detectors with a nonzero amount of dark noise.  As such, the distorted, noisy received signals (images) can result in a degraded SER (the ratio of the number of optical profiles incorrectly classified to the total number of optical profiles received), significantly limiting the ability to implement such systems in a real-world scenario\cite{kaushal2017optical,malik2012influence}.  Here we make use of  generative machine learning techniques to design a communications system that is robust to these real-world hindrances, and demonstrate its ability, in combination with a convolutional neural network, to drastically reduce the effects of turbulence and noise on the SER in a realistic simulated communications setting.

Generative machine learning techniques have recently been developed and applied to a variety of systems, including molecular design, radiotherapy, geophysics, speech recognition and tomography \cite{sanchez2018inverse,jang2018ep,li2018machine,donahue2018exploring,torlai2018neural}. In particular, unsupervised autoencoders have been shown useful in a variety of denoising scenarios \cite{gondara2016medical,fichou2018powerful,cheng2018deep}.  Additionally, several groups have recently investigated the power of convolutional neural networks in the context of optical communications \cite{doster2017machine,park2018multiplexing,lohani2018turbulence,tian2018turbo,li2018joint,lohani2018use,zhao2018mode,jiang2018coherently}.  Here we expand significantly upon these works and develop a generative neural network, combined with a convolutional neural network as shown in Fig.\,\ref{fig:Figure_1}, that act in concert to increase the robustness and decrease the SER and cross-talk of free-space optical communications links.  This receiver-end system is shown to be effective for a wide range of turbulence and detector noise strengths, and requires no feedback to the transmitter of the communications link. Additionally, previous demonstrations have shown that in order to classify unknown distorted optical profiles with high accuracy using only a CNN as a classifier, a training set with large number of known distorted optical profiles is desirable, which is always unbounded. This limits the classification efficiency of the communication scheme with respect to randomly varying turbulent effects. In contrast, our scheme with the developed GML network generates new, significantly less distorted optical profiles at the receiver which is later classified by a CNN exclusively trained with undistorted (desired) optical modes with added dark noise alone. As GML is based on unsupervised learning, our technique may be easily extended to demodulate more complex optical profiles which are not only difficult to label but also difficult to classify accurately with the current supervised CNN techniques in the future. These new aspects of the current demonstration provide a significant step forward in the realistic implementation of error-correction techniques in free-space optical communications, paving the way toward their robust implementation.

The overall design of the communications system in shown schematically in Fig.\,\ref{fig:Figure_1}.  A laser is simulated to be incident on a spatial light modulator (SLM) with a given phase mask, such that the resultant optical spatial mode profile is in a desired superposition of OAM values ranging from 0 (Gaussian) to $\pm10$ (a petal pattern with 20 lobes).  The mode is then simulated to propagate through turbulence, resulting in a distorted profile, after which detector dark noise is added at the receiver.  This noisy, distorted image is then fed into the generative neural network (GNN), which generates a new mode profile that is fed into a CNN that classifies which mode was sent and received (that is, which letter of the alphabet).  Examples of the distorted, noisy images and GNN generated images are shown in Fig.\,\ref{fig:Figure_2}, for varying degrees of signal-to-noise ratios, turbulence strengths, and communication link distances.  This process is repeated many times for all spatial modes (with random turbulences and noises added), and the SER of the system is calculated and compared to the SER when the neural network detection system is not used.  Additionally, we calculate the cross-talk between the noisy, distorted modes at the receiver, and show a significant enhancement when the generative neural network system is used.

\section{Free-Space Turbulent Propagation and Network Architecture}
We describe here in detail the various steps to reconstruct the distorted optical profiles that enhance the SER using the generative denoising autoencoder. We focus on the effects of turbulent propagation and dark noise that seriously degrade the OAM mode quality at the receiver.

First, we simulate the free-space propagation of a Gaussian beam $G(x,y,w_0)$ with waist $w_0$ through a SLM loaded with a phase mask $\Theta^{(\ell_1,\ell_2)}(x,y)$, corresponding to the superposition of two Laguerre-Gauss modes of OAM azimuthal quantum numbers $\ell_1$ and $\ell_2$. This results in intensity profiles at the receiver, a distance $Z\,$m away from the transmitter, that are in the desired optical mode. In order to generate the superposition phase mask at the SLM we use equation (\ref{phase}),
\begin{equation}
\Theta^{(\ell_1,\ell_2)}(x,y) = \angle [\exp(-i\ell_1\angle(x+iy)) + \gamma(r,\ell_1,\ell_2) \exp(-i\ell_2\angle(x+iy))]
\label{phase}
\end{equation}
where ``$\angle$'' represents the arctan of the ratio of the imaginary part to the real part, $\gamma(r,\ell_1,\ell_2) = \sqrt{\frac{|\ell_1|!}{|\ell_2|!}}(\frac{r\sqrt{2}}{w_0})^{|\ell_2| - |\ell_1|}$ with $r$ as the radial distance from the center axis of the beam. Finally, the intensity at the receiver, $I_r$, is found by using the Fourier propagator given by equation (\ref{eqn:Ir}),
\begin{equation}
I_r = \Bigg|\mathcal{F}^{-1}\bigg\{ H \times \mathcal{F} \big[G(x,y,w_0)\,\exp(i\,\Theta^{(\ell,-\ell)})\big]\bigg\}\Bigg|^2
\label{eqn:Ir}
\end{equation}
where  $\mathcal{F}$ is the fast Fourier transformation, and $H$ is the transfer function.

We use a Kolmogorov phase with the Von Karman spectrum effects model\cite{bos2015anisotropic} to simulate the turbulence in the atmosphere, which is given by equation (\ref{eqn:phi_k}),
\begin{equation}
\Phi(\kappa)\,=\,0.023r_0^{-5/3}(\kappa^2 + \kappa_0^2)^{-11/6}\exp{(-\kappa^2/\kappa_m^2)} 
\label{eqn:phi_k}
\end{equation}
where $r_0$\,=\,$(0.423k^2 C_n^2 Z)^{-3/5}$ is the Fried parameter for a propagation distance Z, and k\,=\,$2\pi/\lambda$ is the wave-vector for a given wavelength $\lambda$ of light. Here $\kappa$ is the spatial frequency, $\kappa_m$\,=\,5.92/$l_{min}$, and $\kappa_0$\,=\, $2\pi/l_{max}$, with inner $(l_{min})$ and outer $(l_{max})$ scales of turbulence.  
Finally, we generate random phase screens using the inverse Fourier transformation as given in equation (\ref{eqn:phi_xy}), 
\begin{equation}
\Phi (x,y) = \Re \Big\{\mathcal{F}^{-1}\Big(\mathbb{C}_{NN}\sqrt{\Phi_{NN}(\kappa)}\Big)\Big\}
\label{eqn:phi_xy}
\end{equation}
where $\Re$ means taking only the real part, $\mathcal{F}^{-1}$ is the inverse fast Fourier transformation, $\mathbb{C}_{NN}$ is a complex random normal number with zero mean and unit variance, and $\sqrt{\Phi_{NN}(\kappa)}$ is the square root of the phase distributions given by equation (\ref{eqn:phi_k}) over the sampling grid of size N\,$\times$\,N.

The turbulent environment is simulated by placing a turbulent phase screen generated by equation (\ref{eqn:phi_xy}) a meter away from the SLM plane. 
Then, we propagate the Gaussian beam, $G(x,y,w_0)$, through all the phase screens (SLM and turbulence) and have the intensity profile, $I_r^{t}$, with additive dark noise, $N(0,\sigma)$, at the receiver, $Z \,\geq\,200\,m$ away from the turbulence plane, using equation (\ref{eqn:I_r_t}),
\begin{equation}
I_r^t = \Bigg|\mathcal{F}^{-1}\bigg\{H_2 \times \mathcal{F}\bigg(\mathcal{F}^{-1}\big\{ H_1 \times \mathcal{F} [G(x,y,w_0)\,
\exp(i\,\Theta^{(\ell,-\ell)})]\big\} \exp(i\,\Phi (x,y))\bigg)\bigg\}\Bigg|^2+ N(0,\sigma)
\label{eqn:I_r_t}
\end{equation}
where $H_1$ and $H_2$ are again transfer functions from the SLM to turbulence plane, and the turbulence plane to receiver, respectively. In order to generate the turbulence discussed here, we use $w_0\,=\,4\,$cm, $N\,=\,128$, $\lambda\, = \,1550\,$nm, $l_{min}\,=\,1\,$mm, $l_{max}\,=\,200\,$m, and $C_n^2$ varies from $9 \times 10^{-15}\,m^{-2/3}$ to $5 \times 10^{-13}\,m^{-2/3}$.

\subsection{Generative convolutional denoising autoencoder}
The distorted received optical profiles, $I_r^{t}$, as given by equation (\ref{eqn:I_r_t}) are fed to the encoder of the GNN which compresses them into a latent space S as expressed in equation (\ref{eqn:encoder}),
\begin{equation}
Encoder:\, S\,=\, f_{\theta}(I_r^{t})\,=\,\Big\{\Big(\text{Max}\big( I_r^{t}*w_1^k+b_1^k\big)\Big)
*w_2^k+b_2^k\Big\}\times W+B , \quad  \theta \in \{ w_1^k,\,b_1^k,\,w_2^k,\,b_2^k,\,W,\,B\}
\label{eqn:encoder}
\end{equation}
where $\theta$ represents a parameter space of $w_1^k$, $w_2^k$ (weights), and $b_1^k$, $b_2^k$ (biases) of $k^{th}$ feature mappings of the first and second convolutional layers, respectively, where ``Max'' corresponds to a max-pooling operation. Also, \textit{W} and \textit{B} represent the weight and bias of a fully connected layer, and for convenience $``*"$ represents the convolutional/transpose-convolutional operation. Note that we apply the ReLU activation after each convolutional operation. The resulting latent space S is then forwarded to the decoder that maps it to reconstructed pixels $I$ of the input space as given by equation (\ref{eqn:decoder})
\begin{equation}
Decoder:\, I\,=\, g_{\theta^\prime}(S) =  \Big\{\big((S\times W^\prime+B^\prime)*{w^\prime}_1^n
+{b^\prime}_1^n\big)*{w^\prime}_2^n+{b^\prime}_2^n\Big\}*{w^\prime}+{b^\prime} ,
\quad  \theta^\prime \in \{ {w^\prime}_1^n,\,{b^\prime}_1^n,\,{w^\prime}_2^n,\,{b^\prime}_2^n,\,w^\prime,\,b^\prime,\,W^\prime, \,B^\prime \}
\label{eqn:decoder}
\end{equation}
where primes represent the parameter space of the decoder corresponding to $n^{th}$ feature mappings. Here each training optical profile ${I_r^{t}}^{(i)}$ is successively mapped into a corresponding latent space $S^{(i)}$, and a generation $I^{(i)}$. After that a square reconstruction loss \textbf{L}$(I^{(i)},I_r^{(i)})$ is evaluated, where $I_r^{(i)}$ is an undistorted optical profile at the receiver given by equation (\ref{eqn:Ir}). Finally, in order to optimize the parameters, we minimize the average reconstruction loss given by equation (\ref{eqn:loss}) using adamoptimizer of tensorflow\cite{tensorflow2015-whitepaper}.
\begin{equation}
\theta\,,\theta^\prime = \underset{\theta\,,\theta^\prime}{\text{argmin}}\, \frac{1}{N}\sum_i^N \textbf{L}\big(I^{(i)},\,I_r^{(i)}\big)\,
=\,\underset{\theta\,,\theta^\prime}{\text{argmin}}\,\frac{1}{N}\sum_i^N \textbf{L}\Big(g_{\theta^\prime}\big(f_\theta(I_r^t{^{(i)}})\big),\,I_r^{(i)}\Big)
\label{eqn:loss}
\end{equation}

\subsection{Convolutional neural network as a demodulator}
In order to train a CNN to classify the generated modes, we simulate optical modes at the receiver without any turbulence using equation (\ref{eqn:Ir}) for each OAM superposition value ranging from $\ell\,=\,0$ to $\pm 10$. We then manually add random Gaussian noise with $\sigma\,=\,2$. Here we keep low noise in the training and testing sets of the CNN to estimate how closely the generated modes by the GNN fit with the target mode. Finally we simulate 150 noisy images for each value of OAM for a total of 16,500 images. The image set is split into a training set with 130 images and a test set with 20 images, again, for each OAM profile. Then, the parameter space of the CNN is optimized by minimizing a softmax cross-entropy loss using adamoptimizer. Note that pre-trained CNN network has an unity accuracy with respect to the test images\cite{lohani2018use}. 

\section{Results}
Apart from traditional autoencoder as described in\cite{MAL-006},
 convolutional denoising autoencoders (CDAEs) are able to reconstruct a clean, corrected input from those that are partially distorted \cite{vincent2008extracting}. The idea behind using such a network design is to learn a hidden representation and extract the important features which are robust to noise or distortion present in the inputs. The generative network used here consists of three layers -- an encoder, latent space, and a decoder as shown in Fig. \ref{fig:Figure_2} (a-top). The encoder extracts the important features and compresses them into a smaller size, which is a latent space. The encoded information in latent space is then forwarded to the decoder, which finally generates the desired clean modes. Our CDAE is built with convolutional layers to encode the noisy inputs to latent space and  a transpose-convolutional layer as a generator to decode the latent space. The encoder contains two convolutional layers (green blocks in Fig. \ref{fig:Figure_2} (a)) with a kernel size of $5\times5$, with zero padding, ReLU activation, stride length of 2, and 3 feature mappings followed by a max-pooling layer (red block) with a pool size of $2\times2$, and a single fully connected layer (blue circles) to a latent space. The decoder  begins with a fully connected layer (blue circles) followed by a convolutional layer with the same parameter settings as described above. In order to regain the original size of the input, a transpose-convolutional layer (magenta block) is applied, again with the same parameter values. Finally, a convolutional block with a single feature mapping generates a clean, corrected mode profile. Note that we apply a dropout with a rate of $5\%$ after each layer, except the fully connected layer at the end of the encoder and final convolutional layer of the decoder. We apply the small dropout rate to avoid overfitting, as well as the possible loss of features extracted from the convolution. The size of the fully connected layer is same as that of the latent space. 
Similarly, we implement a CNN to demodulate the generated, reconstructed clean mode profiles as well as the uncorrected received profiles.  This network consists of a single convolutional unit with a kernel of size $5\times5$, zero padding, ReLU activation, stride length of 2, and a single feature mapping followed by a max-pooling with a $2\times2$ filter attached to a fully connected layer ($28\times28$ neurons) and an output layer as shown in Fig. \ref{fig:Figure_2} (a-bottom). No dropout is employed in this network. 
\begin{figure}[h!] 
\centering\includegraphics[width=\linewidth]{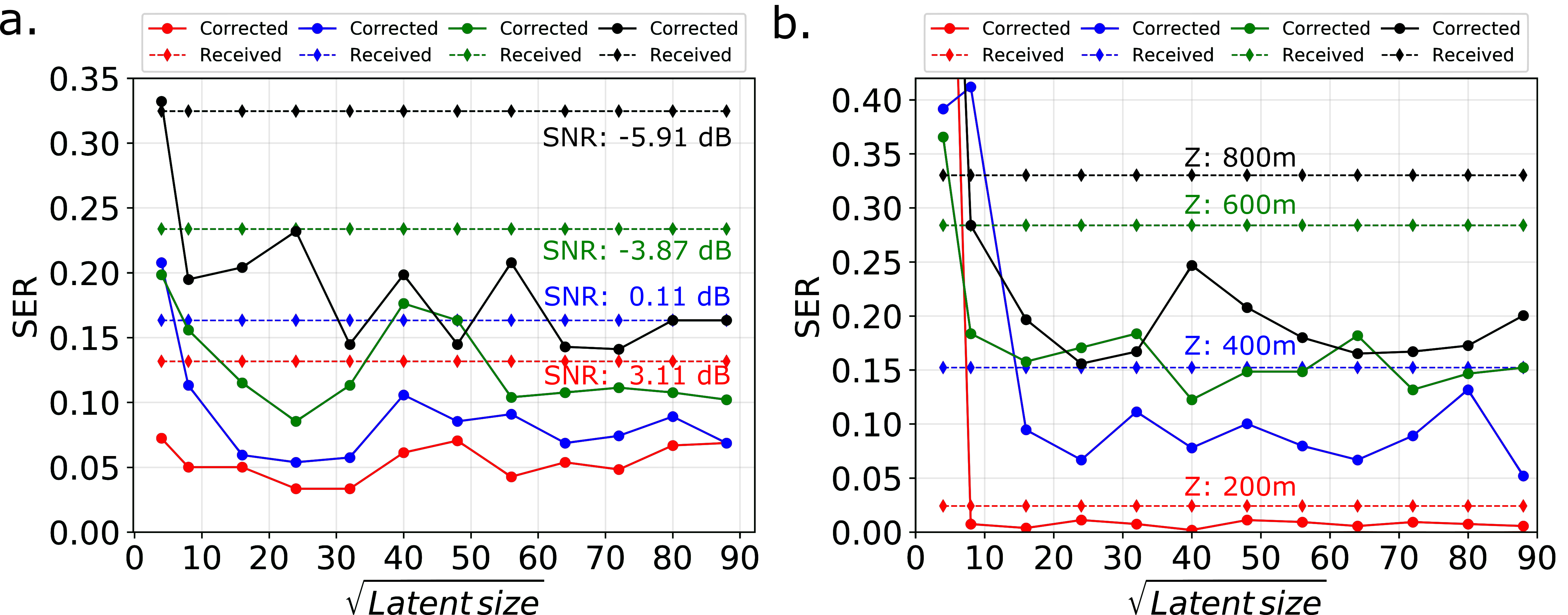}
\caption{SER versus latent space size (a) at different levels of SNR with $C_n^2\,=\,5\times 10^{-14}\,m^{-2/3}$, and $Z\,=\,500\,m$ (b) at various communication links distance with $C_n^2\,=\,5\times 10^{-14}\,m^{-2/3}$, and SNR $=\,-3.87\,$dB, for the received (noisy and distorted) and corrected (generated) mode profiles.} 
\label{fig:Figure_3}
\end{figure}

First we evaluate the SER improvement with respect to latent space size of the GNN for various signal to noise ratios (SNRs) of the received OAM profiles. In order to generate training sets we simulate 99 random turbulent phase screens with a strength $C_n^2$ of $5\times 10^{-14}\,m^{-2/3}$, and a communication link distance (\textit{Z}) of $500\,$m. Note that the 99 simulated phase screens are all different from one another with respect to their phase distributions, such that two different turbulent phase screens produce different scintillation effects on OAM mode propagation even if they have the same turbulence strength. As a result we have 99 different distorted optical profiles for each superposition OAM mode ranging from $\ell\,=\,0$ to $\pm\,10$ for a total of 1,089 images. The resolution of the images is fixed to $128\times128$ pixels for all of the simulations performed in this paper. Also, the total intensity (the sum of all pixel values) for the OAM mode images are normalized to 226955 at the transmitter, in order to simulate a transmitter with a fixed transmission intensity. Finally we add random additive Gaussian noise to the distorted, received optical profiles to simulate the effects of dark noise at the receiver. Then, the SNR of the final noisy, distorted optical profiles is measured as discussed in\cite{lohani2018use} . Note that we train the GNN separately for separate SNR image sets (and that the SNR is decreased by increasing the amount of added detector noise, as the transmitter intensity is held constant). Next, the set of images is split into separate training and test sets. As a result, the training set contains 50 images for each value of $\ell$ for a total of 550 images, and the test set contains 49 images for each value of $\ell$ for a total of 539 images. With these training sets the GNN is then pre-trained with a learning hyper-parameter (rate) set at 0.008. Then, unknown test sets are fed to the pre-trained GNN which  generates nearly ideal corrected optical profiles as the output, some examples of which for $C_n^2\,=\,7\times 10^{-14}\,m^{-2/3}$ are shown in Fig. \ref{fig:Figure_2} (a). The left, middle and right columns in Fig. \ref{fig:Figure_2} (a) represent the noisy received modes (R) and reconstructed (C) profiles with a GNN when the average SNR of images are $0.11\,$dB ($\sigma\,=\,20$), $-3.87\,$dB ($\sigma\,=\,50$), and $-5.91\,$dB ($\sigma\,=\,80$), respectively. Next, in order to calculate the SER, the noisy test optical images (without corrections) and reconstructed/generated images from the GNN are forwarded to a pre-trained CNN and the corresponding SERs are measured. The SER of the test set images with and without the GNN at various SNR levels of $3.11\,$dB to $-5.91\,$dB with respect to different latent sizes of the GNN is shown in Fig. \ref{fig:Figure_3} (a). 
We find an improvement in SER from 0.13 to 0.07 for the image sets with SNR\,=\,$3.11 \,$dB even at a small latent size of $4\times4$. As expected we obtain better reconstructions and more improved SERs as we increase the latent size up to $32\times32$, after which it begins to saturate. The reconstructed images shown in Fig. \ref{fig:Figure_2} (a) are from the GNN with a $32\times32$ latent size. Finally we achieve an improvement in SER from 0.13 to 0.03 at a latent size of $24\times24$, 0.16 to 0.05 at a latent size of $24\times24$, 0.23 to 0.08 at a latent size of $24\times24$, and 0.32 to 0.14 at a latent size of $72\times72$ for the image sets with SNR\,=\,$3.11\,$dB, $0.11\,$dB, $-3.87\,$dB, and $-5.91\,$dB, respectively. 

Next, we vary the communication link distance and find the SER improvement with respect to different latent sizes. The same strength of turbulence as described in previous paragraphs is used, but again with different, random phase patterns. Here all optical mode profiles are assumed to be detected with an average SNR$\,=\,-3.87\,$dB at the receiver. Noisy and reconstructed OAM profiles from the pre-trained GNN for distances of 400\,m, 600\,m, and 800\,m are shown in Fig. \ref{fig:Figure_2} (c). The improvement in SER with respect to various latent sizes of the GNN at communication link distances from 200\,m to 800\,m are shown in Fig. \ref{fig:Figure_3} (b). Here, we find significant improvements in the SERs from $2.4\times10^{-2}$ to $1.8\times10^{-3}$ at a latent size of $40\times40$, 0.15 to 0.05 at a latent size of $88\times88$, 0.28 to 0.12 at a latent size of $40\times40$, and 0.33 to 0.15 at a latent size of $24\times24$ for the communication links distances of 200\,m, 400\,m, 600\,m, and 800\,m, respectively.

\begin{figure}[h!] 
\centering\includegraphics[width=\linewidth]{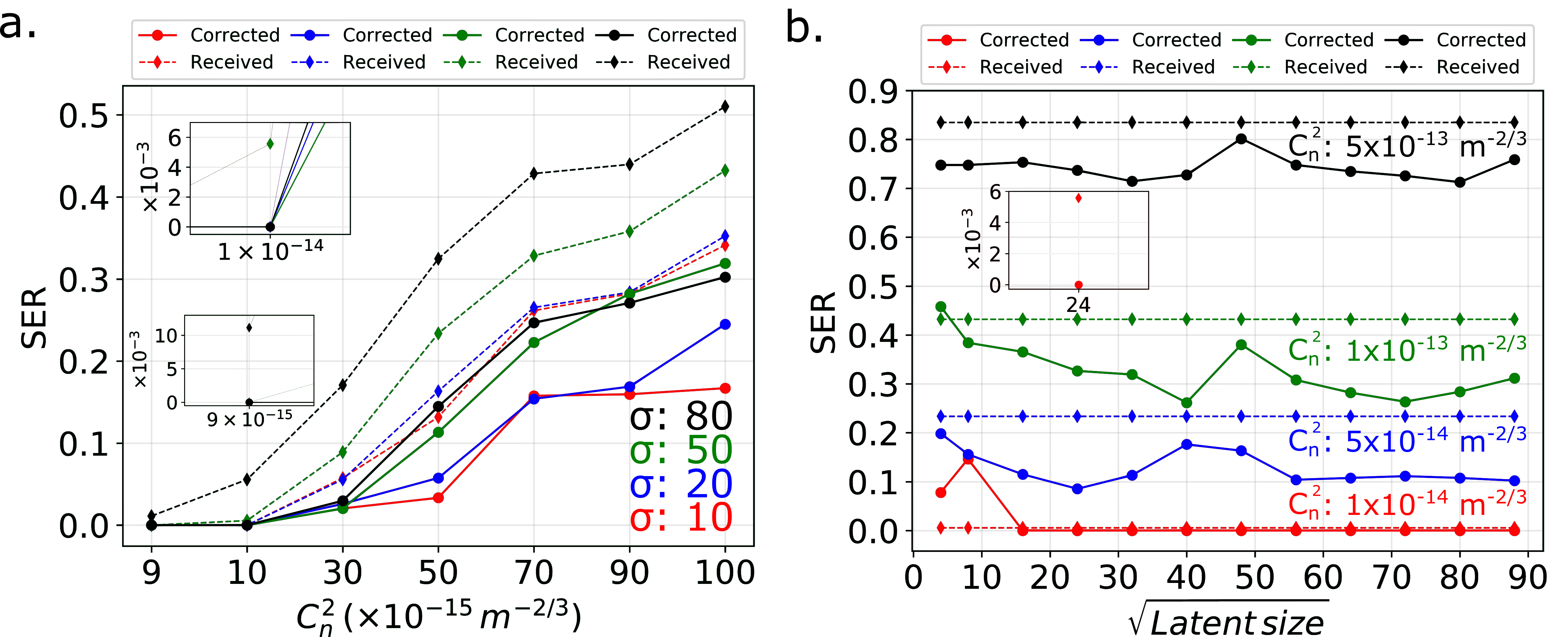}
\caption{(a) SER versus $C_n^2$ with various detector noise levels $\sigma$ at fixed $Z\,=\,500\,m$. The corresponding SER at $C_n^2\,=\,9\times10^{-15}\,m^{-2/3}$, and $C_n^2\,=\,1\times10^{-14}\,m^{-2/3}$ are zoomed in and shown in inset. (b) SER versus latent space size at different turbulent strength $C_n^2$ with fixed $\sigma\,=\,50$, and again $Z\,=\,500\,m$. The inset shows the zoomed in SER at a latent size of $24\times24$.}
\label{fig:Figure_4}
\end{figure}
We now turn to reconstructing severely distorted OAM profiles due to various turbulence strengths in the channel. Turbulence strengths are varied from strong ($C_n^2\,=\,1\times 10^{-13}\,m^{-2/3}$ ) to weak ($C_n^2\,=\,9\times 10^{-15}\,m^{-2/3}$) levels with a fixed $Z\,=\, 500\,$m, for a total of 7 different turbulence strength scenarios, with random phase screens for each class and different noise strengths at the receiver. Some examples of simulated received noisy OAM modes and reconstructed corrected modes using the GNN with a latent size of $32\times32$ for $C_n^2\,=\,5\times 10^{-14}\,m^{-2/3}$, $C_n^2\,=\,7\times 10^{-14}\,m^{-2/3}$, and $C_n^2\,=\,1\times 10^{-13}\,m^{-2/3}$ at a dark noise strength of $\sigma\,=\,50$ are, respectively, shown in left, middle and right column of Fig. \ref{fig:Figure_2} (b). Again with the pre-trained GNN with a latent size of $32\times32$, we show a significant improvement in SERs for various turbulence strengths for noise strengths of $\sigma\,=\,10$, $\sigma\,=\,20$, $\sigma\,=\,50$, and $\sigma\,=\,80$ present at the receiver, as seen in Fig. \ref{fig:Figure_4} (a). 
Significant reductions in the SER are achieved with the GNN, for example lowering the SER from $0.17$ to $2.9\times10^{-2}$ at a noise level of $\sigma\,=\,80$ and turbulence strength of $C_n^2\,=\,3\times 10^{-14}\,m^{-2/3}$. Furthermore, for an intermediate turbulence strength of $C_n^2\,=\,1\times 10^{-14}\,m^{-2/3}$, we find an uncorrected SER of $5.5\times10^{-3}$, and $5.5\times10^{-2}$, respectively  at $\sigma\,=\,50$, and  $\sigma\,=\,80$, which are dramatically enhanced to $0$ with the GNN as shown by upper inset in Fig. \ref{fig:Figure_4} (a). Additionally, the SER of $1.1\times10^{-2}$ at $\sigma\,=\,80$ for a weak turbulence of $C_n^2\,=\,9\times 10^{-15}\,m^{-2/3}$ has also been corrected to a SER of $0$ with the GNN as shown in the lower inset of Fig. \ref{fig:Figure_4} (a). 
Similarly, in order to have a latent size benchmark with respect to different turbulent strengths, we vary the latent size of the GNN and fix the noise strength at $\sigma\,=50\,$. Again with the GNN pre-trained for various latent sizes, the enhancements in SER with respect to latent sizes are shown in Fig. \ref{fig:Figure_4} (b). We find an improvement in SER from $5.5\times10^{-3}$ to $0$ for every latent size at or above $16\times16$ for $C_n^2\,=\,1\times 10^{-14}\,m^{-2/3}$ (red curve). For example, the SER at a latent size of $24\times24$ are zoomed in and shown in the inset. Moreover, we show a significant SER enhancement from $0.23$ to $0.08$ at a latent size of $24\times24$ (blue curve), $0.43$ to $0.26$ at a latent size of $40\times40$ (green curve), and $0.83$ to $0.71$ at latent size of $80\times80$ (black curve), respectively, for an intermediate turbulence strength of $C_n^2\,=\,5\times 10^{-14}\,m^{-2/3}$, strong strength of $C_n^2\,=\,1\times 10^{-13}\,m^{-2/3}$, and extremely strong strength of $C_n^2\,=\,5\times 10^{-13}\,m^{-2/3}$ in Fig. \ref{fig:Figure_4} (b). 
\begin{figure}[h!] 
\centering\includegraphics[width=\linewidth]{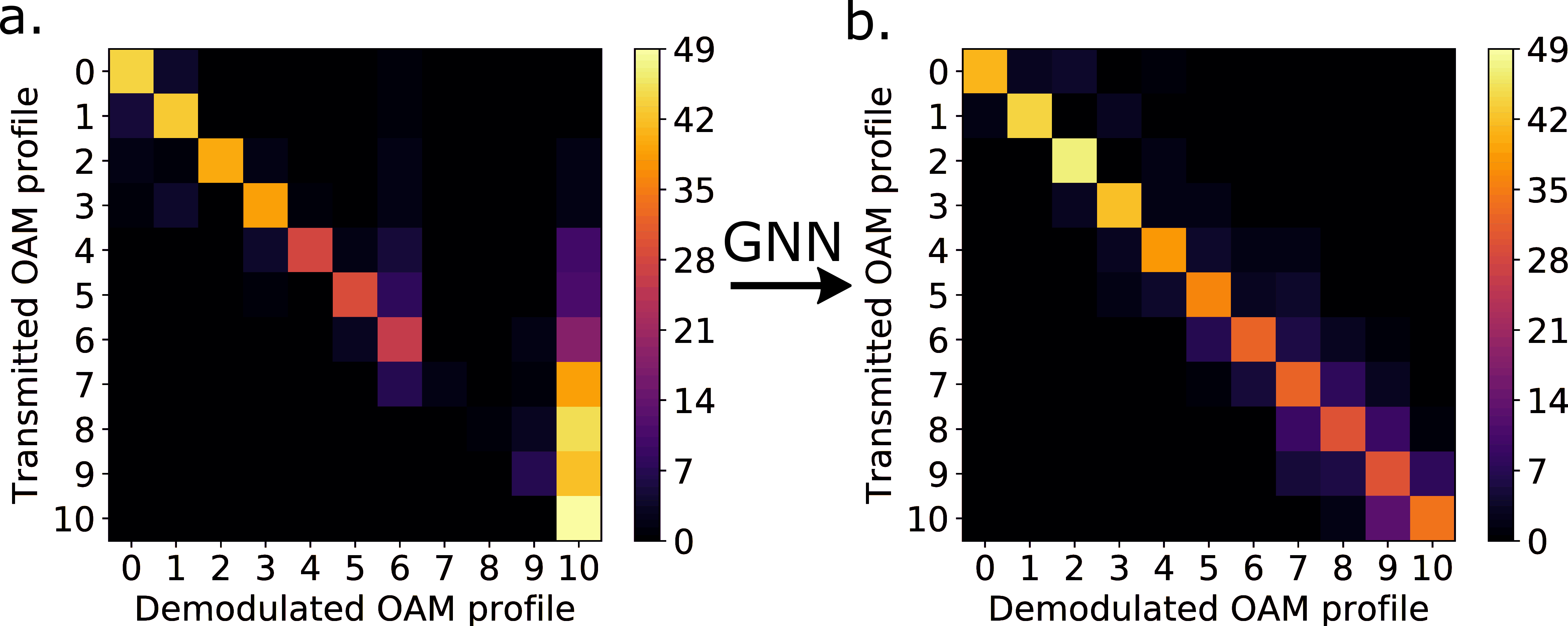}
\caption{Cross-talk states between transmitted and demodulated OAM profiles with $C_n^2\,=\,7\times 10^{-14}\,m^{-2/3}$, $\sigma\,=\,80$, and communication distance of $500\,m$ (a) before applying GNN, and (b) after making correction with GNN.} 
\label{fig:Figure_5}
\end{figure}

Lastly, we investigate improving the SER for individual superposition OAM profiles. In order to illustrate the robustness of the GNN toward the reconstruction of individual OAM profiles to decrease the possible cross-talk with other modes, we take a test set simulated with moderately strong turbulence of $C_n^2\,=\,7\times 10^{-14}\,m^{-2/3}$, a noise strength of $\sigma\,=\,80$, and communication link distance of $Z\,=\,500\,m$. As discussed earlier the test set again contains 49 distorted optical images for each OAM value of $\ell$ ranging from 0 to $\pm 10$. With the pre-trained GNN with a latent size of $32\times32$, we reconstruct 49 OAM images for each OAM mode for the given test set and predict the corresponding mode value using the CNN. The mode values predicted by the CNN without and with the GNN versus the correct transmitted OAM profile are shown in the cross-talk plot of Fig. \ref{fig:Figure_5} (a) and (b), respectively. We find the majority of the uncorrected optical profiles with OAM mode values $\ell\,=\,\pm6,\,\pm7,\,\pm8,\,\pm9$ are incorrectly demodulated at the receiver as shown in Fig. \ref{fig:Figure_5} (a). However, with the GNN reconstruction of the modes, the incorrect predictions have been significantly reduced from $23$ to $17$, $47$ to $17$, $48$ to $19$, and $42$ to $19$ out of $49$ test images for each OAM mode value of $\ell\,=\,\pm6,\,\pm7,\,\pm8,\,\pm9$, respectively, as shown in Fig. \ref{fig:Figure_5} (b).

\section{Discussion}
In conclusion, we have developed an efficient and straightforward scheme to overcome the negative effects of turbulence and detector noise in FSO communications with a GML and CNN in combination. The developed state-of-the-art technique corrects for distortions caused by weak to extreme strengths of turbulence, as well as various strengths of detection noise, resulting in significantly improved SERs at the receiver. We also show an enhancement in SER for various communication link distances. Additionally, by using the same network system, we demonstrate the robustness of the GML approach with respect to the reconstruction of individual OAM profiles, which results in a decrease in the cross-talk between received mode profiles. Moreover, with the aid of generative networks, we have significantly improved SERs with a CNN as a classifier that is solely trained with undistorted modes with dark noise at the receiver, thereby avoiding the need for extremely large CNN training sets involving various distortions (which is unbounded). Furthermore, the addition of this unsupervised learning scheme may be extended to demodulate more complex optical profiles which are difficult to label and classify with current supervised techniques. This significant improvement in mode classification and demodulation is integral to the robust performance of realistic FSO communications systems, and we are hopeful that the techniques developed here may directly be applied to quantum systems in the near future \cite{hughes2002,paterson2005,swaim2017,tyler2009,gupta2016,krenn2014}.

\bibliography{autoencoder}

\section*{Acknowledgements}
We acknowledge funding from the U.S. Office of Naval Research under grant number N000141912374.

\section*{Author contributions statement}
S.L. developed the neural network system, and ran all simulations.  R.T.G. developed the project.  Both authors prepared the manuscript.

\end{document}